\documentclass{article}
\usepackage{spconf_Amir}
\usepackage{amsmath,graphicx}
\usepackage {algorithm,algorithmic}
\usepackage{color,soul}
\usepackage{fancyhdr}

\title{2.5D Deep Learning for CT Image Reconstruction using a Multi-GPU implementation}
\name{Amirkoushyar Ziabari$^{\star}$, Dong Hye Ye $^{\star}$ $ ^{\dagger}$, Somesh Srivastava$^{\ddagger}$, Ken D. Sauer $^{\oplus}$}
\secondlinename{Jean-Baptiste Thibault $^{\ddagger}$, Charles A. Bouman$^{\star}$ }%
\address{$^{\star}$ Electrical and Computer Engineering at Purdue University \\
     $^{\dagger}$ Electrical and Computer Engineering at Marquett University \\
     $^{\ddagger}$ GE Healthcare  \\ 
     $^{\oplus}$ Electrical Engineering at University of Notre Dame}
\begin{document}

\maketitle
%

\begin{abstract}
While Model Based Iterative Reconstruction (MBIR) of CT scans has been shown to have better image quality than Filtered Back Projection (FBP),
its use has been limited by its high computational cost.
More recently, deep convolutional neural networks (CNN) have shown great promise in both denoising and reconstruction applications. 
In this research, we propose a fast reconstruction algorithm, which we call Deep Learning MBIR (DL-MBIR),
for approximating MBIR using a deep residual neural network.
The DL-MBIR method is trained to produce reconstructions that approximate true MBIR images using a 16 layer residual convolutional neural network
implemented on multiple GPUs using Google Tensorflow.
In addition, we propose 2D, 2.5D and 3D variations on the DL-MBIR method and show that the 2.5D method achieves similar quality to the fully 3D method, 
but with reduced computational cost.
\end{abstract}
\section{Introduction}
\label{sec1}

Computed Tomography (CT) plays a vital role in a wide range of applications from diagnosis in health care, to detection in security, and NDE in manufacturing.
While Model Based Iterative Reconstruction (MBIR) has been shown to improve image quality and remove artifacts as compared to more traditional Filtered Back-Projection (FBP) methods~\cite{Ye2017, Yu2011, Thibault2007, Ziabari2018},
a key factor limiting broad acceptance of MBIR is its high computational cost. 

Recently, deep learning approaches have been shown to be of great value in a variety of CT applications. 
Among pioneering examples, Kang, et al. developed a deep convolutional neural network (CNN) 
to remove low-dose noise from X-ray CT images~\cite{Kang2017}.
Excellent performance of deep residual neural networks in image denoising~\cite{He2016, Zhang2017a}, inspired Han, et al.\ to propose a residual learning method to remove the global streaking artifacts caused by sparse projection views in CT images~\cite{han2016deep}.
More recently, Ye, et al.~\cite{ye2018deep} developed method for incorporating CNN denoisers into MBIR reconstruction as advanced prior models using
the Plug-and-Play framework \cite{venkatakrishnan2013plug, sreehari2016plug}.

In this paper, we propose a fast reconstruction algorithm, which we call Deep Learning MBIR (DL-MBIR),
for approximately achieving the improved quality of MBIR using a deep residual neural network.
The DL-MBIR method is trained to produce 3D reconstructions that approximate true MBIR images using a 16 layer residual convolutional neural network implemented on multiple GPUs using Google Tensorflow.
We present three implementations of DL-MBIR corresponding to processing the data in 2D, 2.5D and 3D.
While the 3D processing is shown to offer the best fidelity to MBIR reconstruction, it requires 3D convolutions that increase computation relative to the 2D approach.
Alternatively, the 2.5D method results in image fidelity that is comparable to 3D processing;
but since it only requires a 2D convolution structure, its computational requirements are similar to 2D processing.
Finally, we present results on clinical case studies that demonstrate value of DL-MBIR in general and the favorable speed/fidelity trade-off 2.5D processing.

\section{2D and 3D Deep Learning MBIR (DL-MBIR)}
\label{sec2}

The architectural structure of DL-MBIR is shown in Figure~\ref{fig1}.
The aim of this network is to find a nonlinear mapping that transforms the FBP image into an accurate approximation of the MBIR image.


\subsection{2D DL-MBIR}

\begin{figure}[ht]
\centering
\includegraphics [scale=.34]{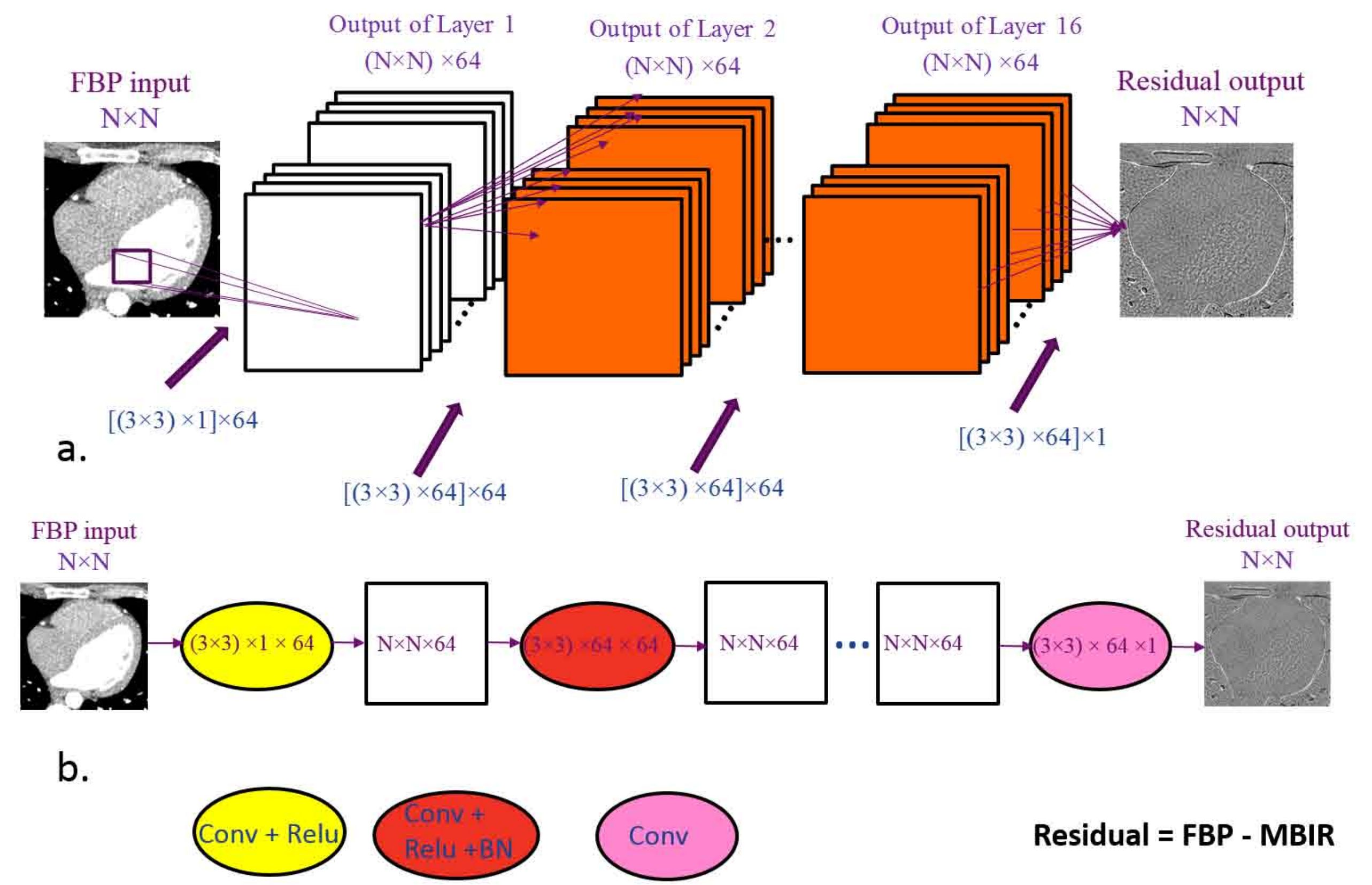}
\caption{Architecture of 2D DL-MBIR network. 
a. Detailed, and b. simple view. 
The network consists of 17 layers.
Importantly, the input is a single 2D slice from the FBP image. }
\label{fig1}
\end{figure}

The DL-MBIR network shown in Figure~\ref{fig1} is called 2D DL-MBIR because it processes 2D input slices and only uses 2D convolutional kernels.
This network consists of 17 layers, and for each layer we represent the transformation kernel in the form 
$$
(3 \times 3) \times N_i \times N_o \ ,
$$
where $(3 \times 3)$ represents the convolution kernel applied with $N_i$ input channels and $N_o$ output channels.
In the first layer, we apply 64 convolutional filter kernels each of size $(3\times3)$ followed by a rectified linear units (ReLU)~\cite{glorot2011deep}
to form 64 output channels for the first layer.
This kernel is denoted as $(3\times 3) \times 1 \times 64$ since it has only a single slice of an FBP image as input.
From layers 2 to 16, we a apply a kernel of size $(3 \times 3)\times 64 \times 64$ as to generate a 64 channel out form a 64 channel input from the previous layer.
Between each convolution and ReLU, we added batch normalization for better convergence and stability of the network~\cite{ye2018deep, ioffe2015batch}.
In the last layer, we apply a $(3\times 3) \times 64 \times 1$ kernel to generate a single output residual image.

In this network, and in general in DL-MBIR networks, we use the residual learning strategy~\cite{He2016, Zhang2017a}, where the network is trained on pairs of FBP/residual image patches. 
The residual corresponds to the difference between the MBIR and FBP reconstructions of the same scan data.
Assuming $y \in \Re^N$ and $x \in \Re^N$ are FBP and MBIR data, the residual training images $\nu \in \Re^N$ can be obtained from:

\begin{equation}
   \nu = y - x  
\end{equation}

To find the nonlinear mapping, $R(y;\theta)$, we minimize the following cost function:
\begin{equation}
   \ell(\theta) = \frac{1}{2N_T} \Sigma_{i=1}^{N_T} ||R(y;\theta) - \nu ||^2 \ ,
\end{equation}
where $N_T$ is the total number of training data sets per batch and $\theta$ corresponds to the network parameters.

For robust performance of DL-MBIR, we need to train the deep neural networks on very large data sets that are representative of various clinical settings. 
Since a standard GPU has only 12 or 16 GB memory, multi-GPU implementation is required for deep learning on any large training database. 
This multi-GPU implementation can also reduce the training time through increased parallelization. 

\begin{figure}[ht]
\centering
\includegraphics [scale=.18]{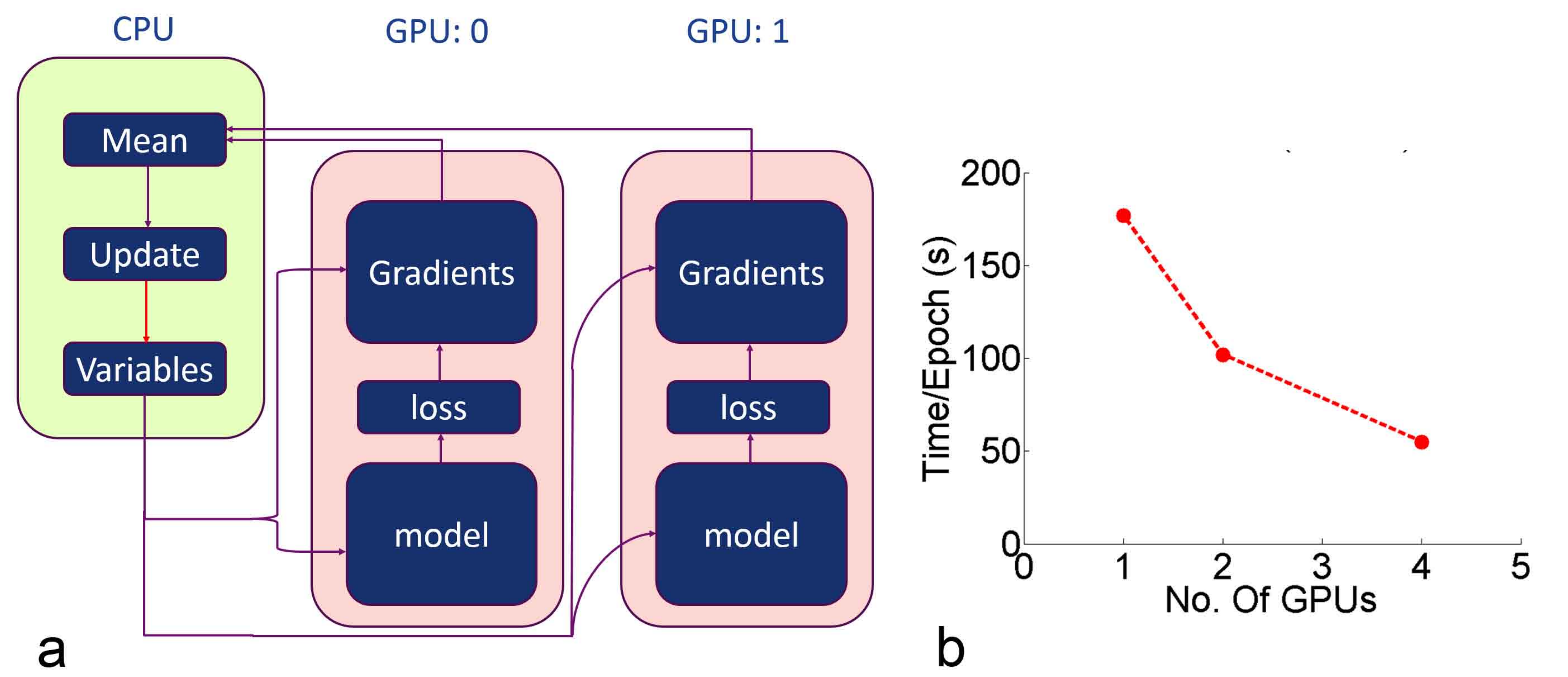}
\caption{a. The strategy for multi-GPU implementation of DL-MBIR on the Google Tensorflow (TF). 
b. Training speed-up with multi-GPU implementation.}
\label{fig2}
\end{figure}

We implemented 2D DL-MBIR in a multi-GPU platform using the standard Google Tensorflow. 
The procedure for multi-GPU implementation is shown in Figure~\ref{fig2}a.
Separate GPUs calculate loss and gradient of a batch of input data. 
After each GPU completes the calculation of gradients, the gradients are averaged on the CPU (for synchronization of the data), and the loss is minimized using the ADAM optimizer. 
This procedure continues until all the training data are used as input (one epoch).
The training time per epoch is shown in Figure~\ref{fig2}b. 
We used 1, 2 and 4 GPUs on a machine with Titan X (Pascal) GPUs (16G RAM). 

It is worth noting that input batch sizes for 1, 2 and 4 GPUs are 128, 64 and 32, respectively. 
In this manner, the total number of batches processed per iteration are the same for a fair comparison.
Therefore, we can increase the size of training data without sacrificing training time with our multi-GPU TF implementation. 

\subsection{3D DL-MBIR}

\begin{figure}[h]
\centering
\includegraphics [scale=.32]{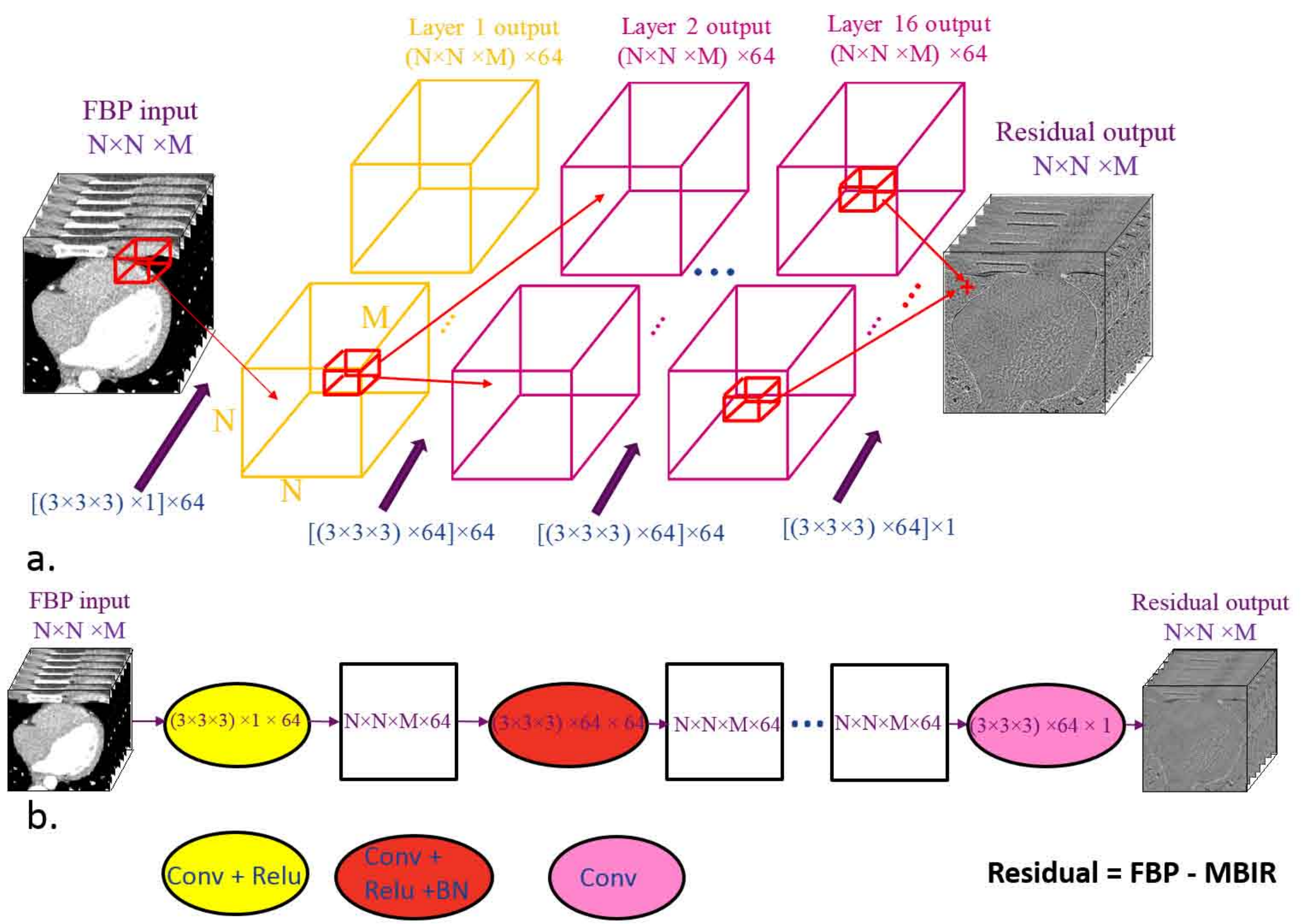}
\caption{Architecture of 3D DL-MBIR network. 
a. Detailed, and b. simple view. 
The input patches and all the convolutions are in 3D.
Seven neighboring slices were used to generate 3D input patches.
Each convolution is represented by (Kernel Size) $\times$ No. of Input Channels $\times$ No. of Filters.}
\label{fig3}
\end{figure}

To improve the performance of DL-MBIR, it is useful to take advantage of of z-direction (depth) information in 3D CT scans.
To that end, we propose to use 3D DL-MBIR instead of 2D DL-MBIR. 
The structure of 3D DL-MBIR network is shown in Figure~\ref{fig3}.

In this network, both the input images and the associated convolution kernels are 3D. 
For the case studies shown in this manuscript, we used convolutions of size (3$\times$3$\times$3) in all cases.
Moreover, for training data, the input patches are generated from 7 neighboring slices and the rest of dimensions and parameters are kept the same between the three networks.
The output volume is also 3D and includes the same number of slices as the input.
For testing, in theory each set of 7 slices as input can results in 7 output slices, but there will be artifact at the edge slices due to truncation. Thus, in the test mode, we used a sliding window of 7 slices that moved with stride 1 in z-direction over the volume of input data, and output only the middle slice from every 7 reconstructed output slices to avoid truncation effect.

\section {2.5D DL-MBIR}
\label{sec3}

In this section, we introduce a 2.5D algorithm in order to maintain the quality of 3D DL-MBIR,
while achieving reconstruction speed similar to 2D DL-MBIR.

\begin{figure}[ht]
\centering
\includegraphics [scale=.32]{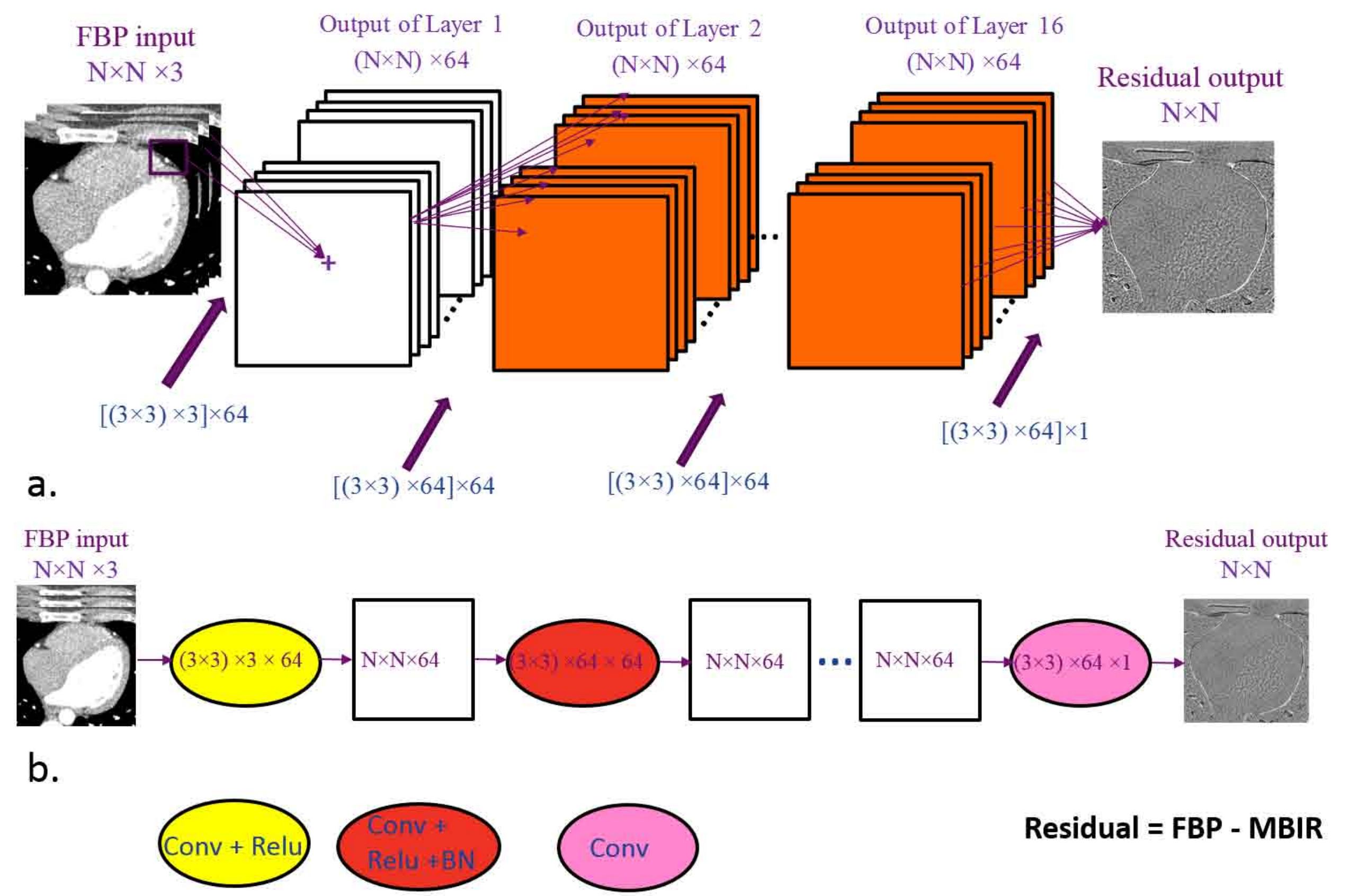}
\caption{Architecture of 2.5D DL-MBIR network. 
a. Detailed, and b. simple view. 
The main difference with Figure~\ref{fig1} (2D DL-MBIR method ) is in the number of input channels (3D vs. 2D input).
Three neighboring slices were used to generate 3D input patches.
Each convolution is represented by (Kernel Size) $\times$ No. of Input Channels $\times$ No. of Filters.}
\label{fig4}
\end{figure}

The structure of the 2.5D DL-MBIR is illustrated in Figure~\ref{fig4}. 
This structure is similar to the 2D DL-MBIR network with convolutional kernels that are 2D, and the kernel size is (3$\times$3). 
However, the 2.5D DL-MBIR takes an input 3 neighboring slices of the FBP reconstruction, but it outputs only a single slice.
The full 3D output is then generated using a sliding window of 3 input slices in z-direction.
In practice more than 3 slices can be used as input without significantly impacting the computational time of the method.
In the results section, we investigate the impact of increasing number of input slices on the 2.5D DL MBIR method.

\section{Results and Discussion}
\label{sec4}

We trained and tested the proposed networks on clinical data sets.
In training mode, we used patches of data to avoid memory problems and also to be able to use more input data~\cite{lee2018Deep}.
In inference mode, however, we used full image slices as input to avoid artifacts due to patching.
Two clinical data sets were used to generate patches of data for training the networks. 
Each data set included FBP and MBIR reconstruction of clinical CT scan data which consisted of 236 slices of 360$\times$360.
Using these data sets along with data augmentation~\cite{ Wong2016}, separate input volumes were created for each method. 
2D DL-MBIR uses 576000 patches of 30$\times$30 for training and while in the 3D DL-MBIR  we used 576000 patches of 30$\times$30$\times$7.
For the 2.5D DL-MBIR we investigated the impact of number of input slices and performed training with data sets with 576000 patches of 30$\times$30$\times$3, 576000 patches of 30$\times$30$\times$5, and 576000 patches of 30$\times$30$\times$7.

We used 80$\%$ of input data for training and 20$\%$ for validation.
Three GPUs were used for training with a batch size of 64 per GPU (192 per iteration).
The trained networks then tested on two other clinical data sets (inference mode).
In both cases, the test data included volumes of FBP and MBIR reconstruction of a CT scan data with 236 slices of 360$\times$360.
The MBIR volume was assumed to be the \emph{ground truth} and the FBP volume was input to the deep learning networks.

The computational time for the test cases are listed in Table~\ref{Tab_compTime}.
The total computational time for testing was 7.4s with the 2D DL-MBIR method.
In the case of 2.5D DL-MBIR, we tested using three different sliding window sizes of 3, 5 and 7.
The computational times were 7.51s, 7.65s, and 7.92s, respectively, which are quite close to the 2D case and slightly different from each other. 
In the case of 3D, however, the test was significantly slower and took 119.4s.

\begin{table}[h]
\vspace{-0.3cm}
\centering
\caption{Testing computational time for DL-MBIR methods }
\label{Tab_compTime}
\resizebox{\columnwidth}{!}{%
\begin{tabular}{|c|c|c|c|c|c|}
\hline
\textbf{Method} & \textbf{2D DL-MBIR} & \textbf{2.5D DL-MBIR } & \textbf{2.5D DL-MBIR} & \textbf{2.5D DL-MBIR} & \textbf{3D DL-MBIR}\\ 
\textbf{} & \textbf{} & \textbf{(3 Slices)} & \textbf{(5 Slices)} & \textbf{(7 Slices)} &  \textbf{} \\
\hline
Computational Time & 7.4s & 7.51s & 7.65s & 7.92s & 119.4s \\ \hline
\end{tabular}
}
\end{table}

\begin{figure}[ht]
\centering
\includegraphics [scale=.17]{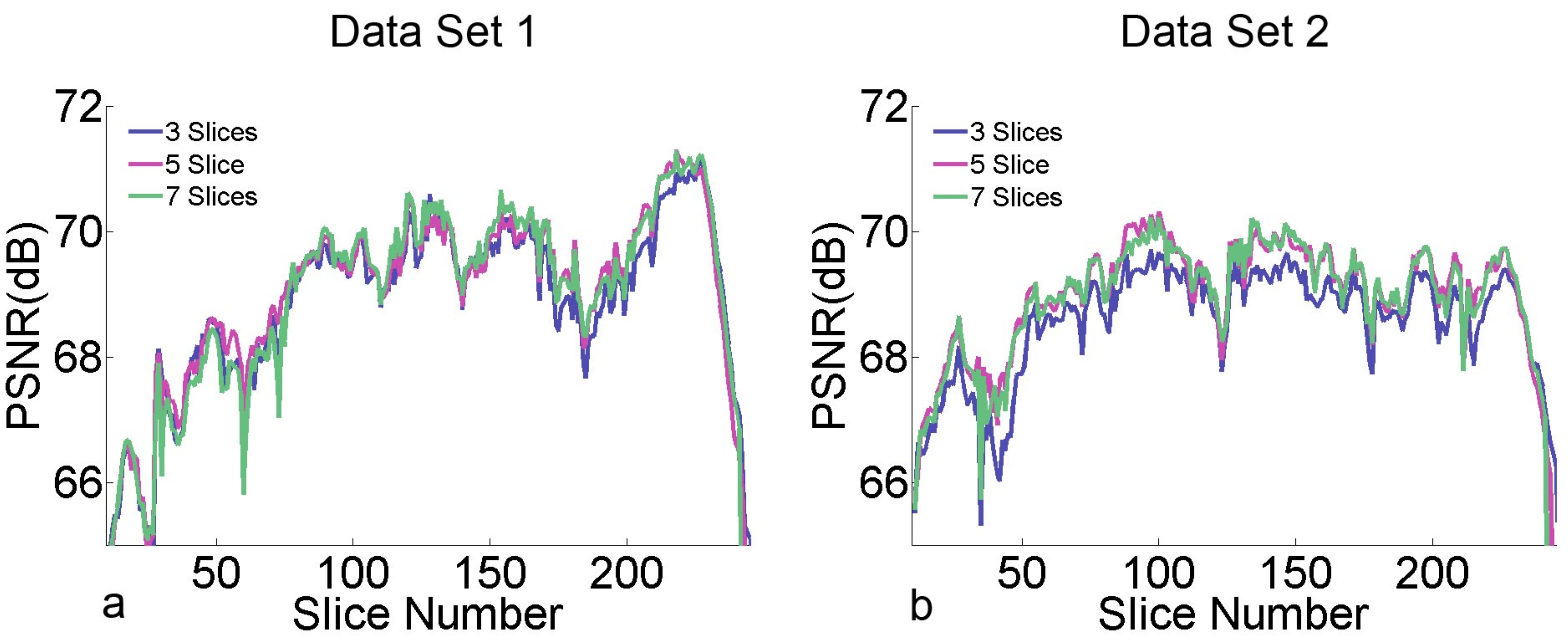}
\vspace{-0.3cm}
\caption{Impact of increasing the number of input slices (sliding window size) on 2.5D DL-MBIR method.
PSNR as a function of slice numbers:
a. Data set 1; and 
b. Data set 2.}
\label{fig5}
\end{figure}

We investigated the impact of increasing the sliding window size on the quality of reconstruction using 2.5D DL-MBIR.
The results for 3 different window sizes of 3, 5 and 7 slices are shown in Figure~\ref{fig5}.
In each case we plotted the Peak-Signal-to-Noise-Ratio (PSNR) as a function of the slice number.

The PSNR was calculated using: 
$$
PSNR \equiv 10 \log \left( \frac{1}{MSE} \right)
$$ 
where we define the Mean Squared Error (MSE) as: 
$$MSE \equiv\frac{1}{N}\sum_{i=1}^{N}(x_{MBIR}(i) - x(i))^2$$
Here, $N$ is the total number of voxels, $x_{MBIR}$ is the voxel that belongs to MBIR volume and $x$ is the corresponding voxel in the reconstructed volume.
To avoid voxels in the air or background of the image slices, we only used voxels that are in the range $[700 ,1500]$~HU in the MBIR images.

Figure~\ref{fig5} a and b correspond to data sets 1 and 2.
In both data sets, using 5 slices as input in the 2.5D DL-MBIR network improved the PSNR compared to the runs with 3 input slices.
The improvement is more prominent in the second data set (Figure~\ref{fig5}b).
A maximum of 1.6~dB improvement in PSNR  in case 2, and a maximum of 0.7~dB improvement in case 1, were observed using 5 input slices instead of 3 input slices. 
On the other hand, using a window size of 7 input slices produced similar results to the runs with 5 input slices.

\begin{table}[h]
\vspace{-0.5cm}
\centering
\caption{Maximum PSNR improvement (compared to the FBP), and average PSNR values in the two test case studies}
\label{Tab1}
\resizebox{\columnwidth}{!}{%
\begin{tabular}{|c|c|c|c|c|}
\hline
\textbf{Method} & \textbf{FBP} & \textbf{2D DL-MBIR} & \textbf{2.5D DL-MBIR} & \textbf{3D DL-MBIR}\\ \hline
Max improvement in dB (Case 1) & - & 4.8 & 5.9 & 6 \\ \hline
Max improvement in dB  (Case 2) & - & 4.5 & 5.1 & 5.5 \\ \hline
Mean PSNR in dB (Case 1) & 64.6 & 68.2 & 68.9 & 68.9 \\ \hline
Mean PSNR in dB (Case 2) & 64.7 & 67.9 & 68.9 & 68.7 \\ \hline
\end{tabular}
}
\end{table}

\begin{figure}[ht]
\centering
\includegraphics [scale=.16]{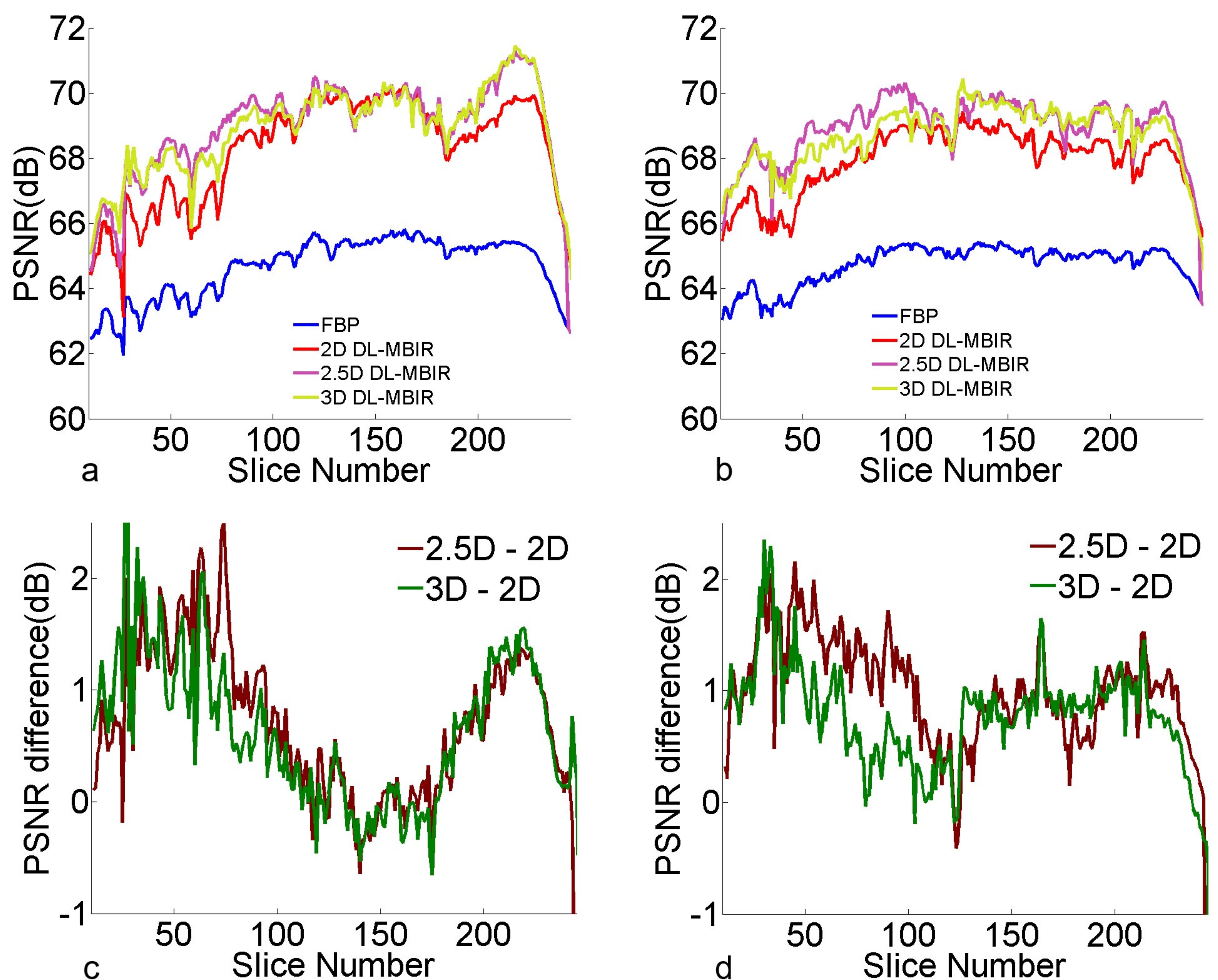}
\vspace{-0.3cm}
\caption{PSNR comparison at each slice between different methods for two different test case studies.
Here, FBP (blue), 2D (red), 2.5D (magenta) and 3D DL-MBIR (green).
a. Data set 1; and 
b. Data set 2.
Difference in PSNR at each slice between 2.5D and 2D (burgundy) and between 3D and 2D (dark green):
c. Data set 1; and 
d. Data set 2.
}
\label{fig6}
\end{figure}

Next, we compared the performance of the proposed 2.5D DL-MBIR method with FBP, 2D and 3D DL-MBIR.
The results are shown in Figure~\ref{fig6}.
The Peak-Signal-to-Noise-Ratio (PSNR) at each slice computed for FBP, 2D, 2.5D and 3D DL-MBIR, which are plotted in Figures 6a and b.
All three DL-MBIR methods produce high quality MBIR-like images and significantly reduce noise and artifacts in comparison with the FBP method.
In addition, in Figures~\ref{fig6}c and d, for both case studies, we plotted the PSNR difference between the 2D and 2.5D DL-MBIR, as well as between 2D and 3D DL-MBIR methods.
It is evident from this figure that the 2.5D DL-MBIR performs on a par with, or in many slices better than, the 3D DL-MBIR, while both methods outperforms the 2D DL-MBIR method.
Both the 2.5D and the 3D DL-MBIR has a maximum PSNR increase of about 2.2~dB with respect to the 2D-DL MBIR method.
More details are provided in Table~\ref{Tab1}.
The maximum PSNR improvement (with respect to the FBP method) and the average PSNR for the 236 slices in each case study are listed in this Table.

Considering the performance of the networks, the better performance of 2.5D DL-MBIR was achieved with negligible computational overhead in comparison with 2D DL-MBIR, and is a promising alternative to pursue in order to improve the reconstruction quality. 

\begin{figure}[ht]
\centering
\includegraphics [scale=.1]{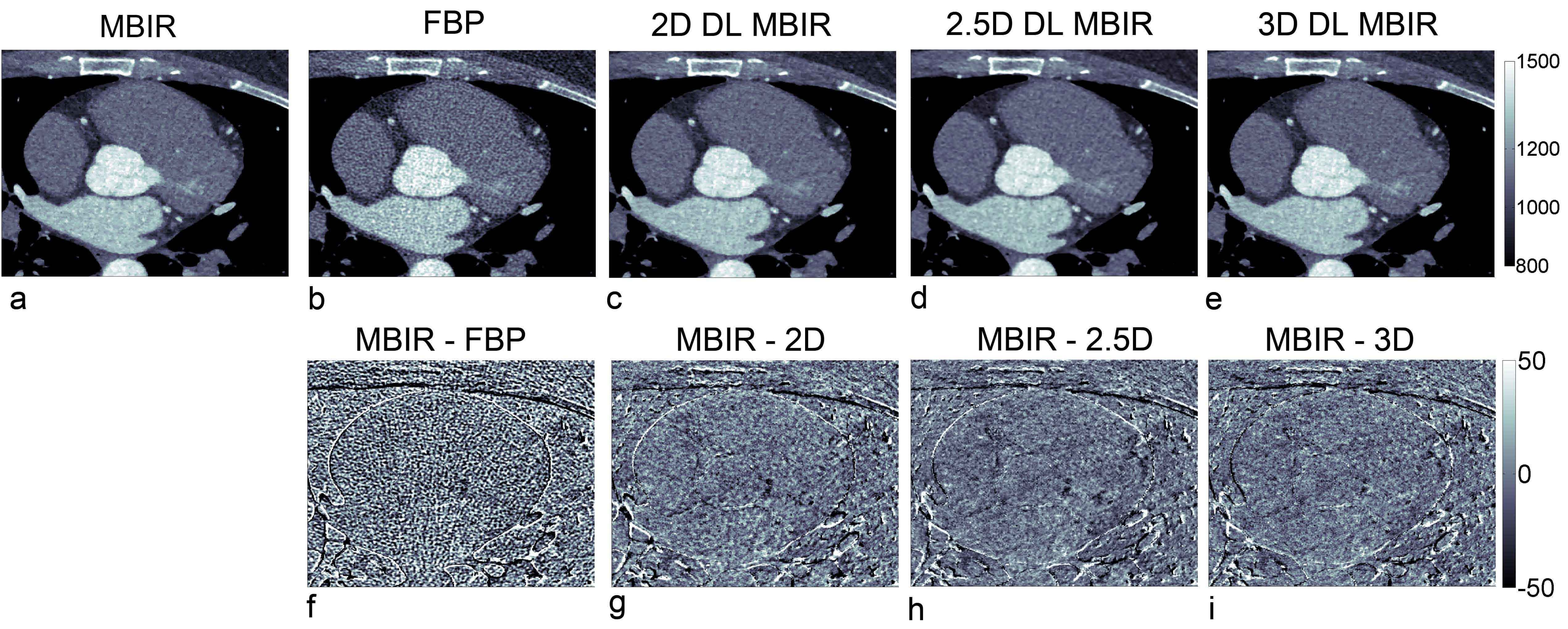}
\vspace{-0.7cm}
\caption{Comparison between different reconstruction of slice 90 in case study 1.
a) MBIR, b) FBP, c) 2D, d) 2.5D, and e) 3D DL-MBIR.
Difference between MBIR and results from: 
f) FBP, g) 2D, h) 2.5D, and i) 3D DL-MBIR.}
\label{fig7}
\end{figure}

\begin{table}[h]
\vspace{-0.5cm}
\centering
\caption{Average PSNR values in the two test case studies}
\label{Tab3}
\resizebox{\columnwidth}{!}{%
\begin{tabular}{|c|c|c|c|c|}
\hline
\textbf{Method} & \textbf{FBP} & \textbf{2D DL-MBIR} & \textbf{2.5D DL-MBIR} & \textbf{3D DL-MBIR}\\ \hline
PSNR (dB) (slice 90, case 1) & 64.8 & 68.3 & 69.93 & 69.51 \\ \hline
PSNR (dB) (slice 150, case 2) & 65.3 & 69 & 70.1 & 69.7 \\ \hline
\end{tabular}
}
\vspace{-0.2cm}
\end{table}

\begin{figure}[ht]
\centering
\includegraphics [scale=.1]{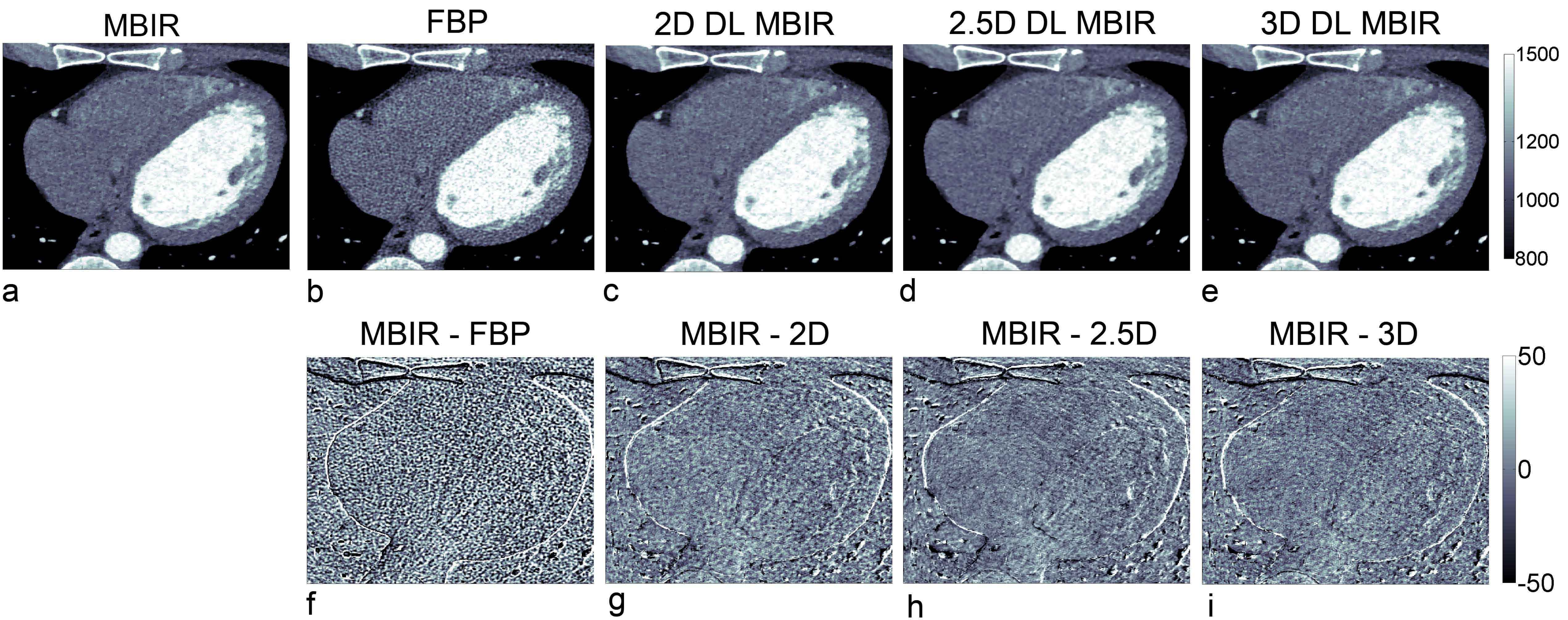}
\vspace{-0.7cm}
\caption{Comparison between different reconstruction of slice 150 in case study 2.
a) MBIR, b) FBP, c) 2D, d) 2.5D, and e) 3D DL-MBIR.
Difference between MBIR and results from: 
f) FBP, g) 2D, h) 2.5D, and i) 3D DL-MBIR.}
\label{fig8}
\end{figure}

\begin{figure}[]
\centering
\includegraphics [scale=.1]{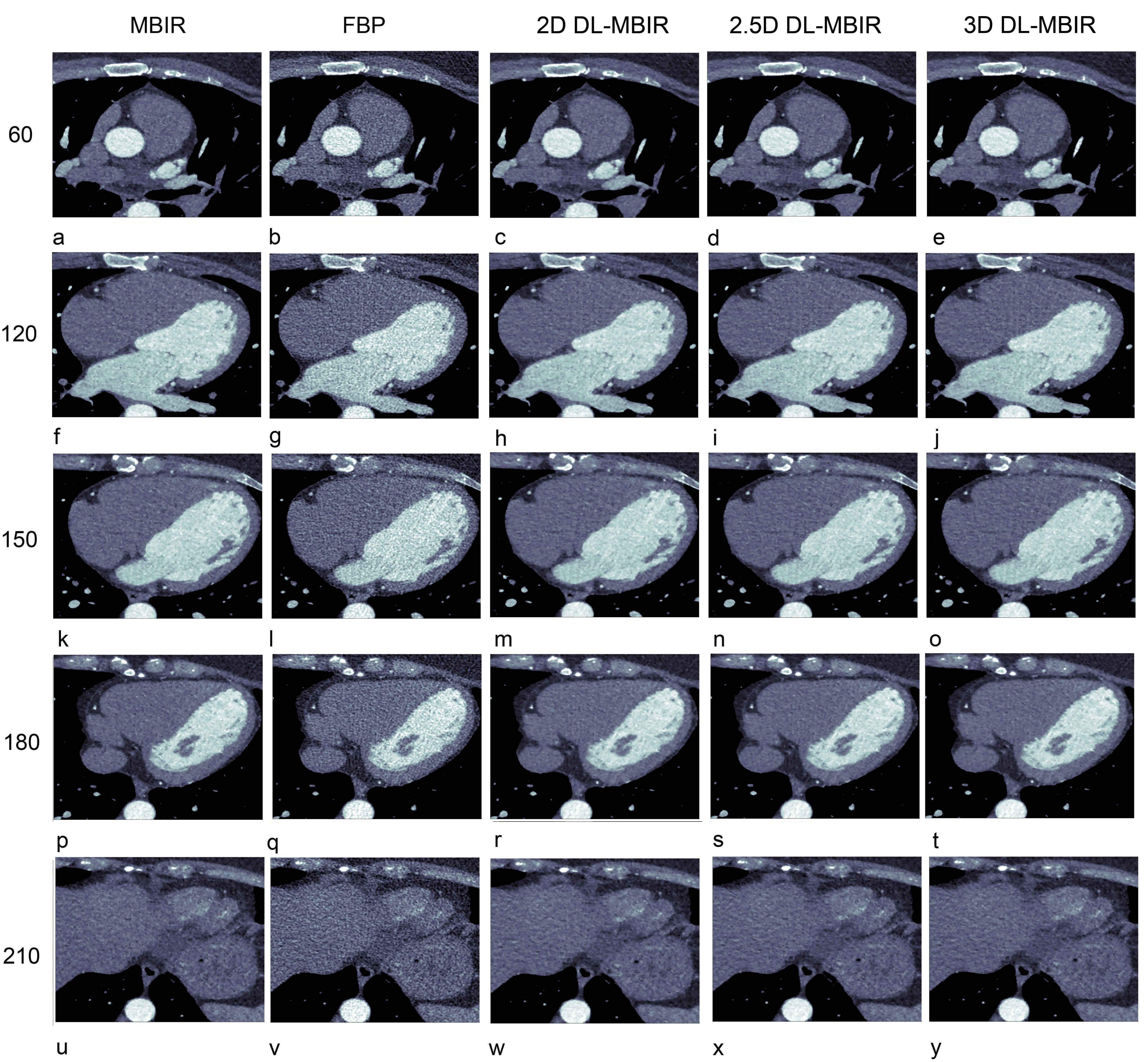}
\vspace{-0.7cm}
\caption{ Comparison between selected slices from different regions of the reconstructed volume.
The slice numbers are:
a-e) 60; f-j) 120; k-o) 150; p-t) 180; u-y) 210.
Columns from left to right: MBIR, FBP, 2D, 2.5D and 3D DL-MBIR.}
\label{fig9}
\end{figure}

\begin{figure}[ht]
\centering
\includegraphics [scale=.1]{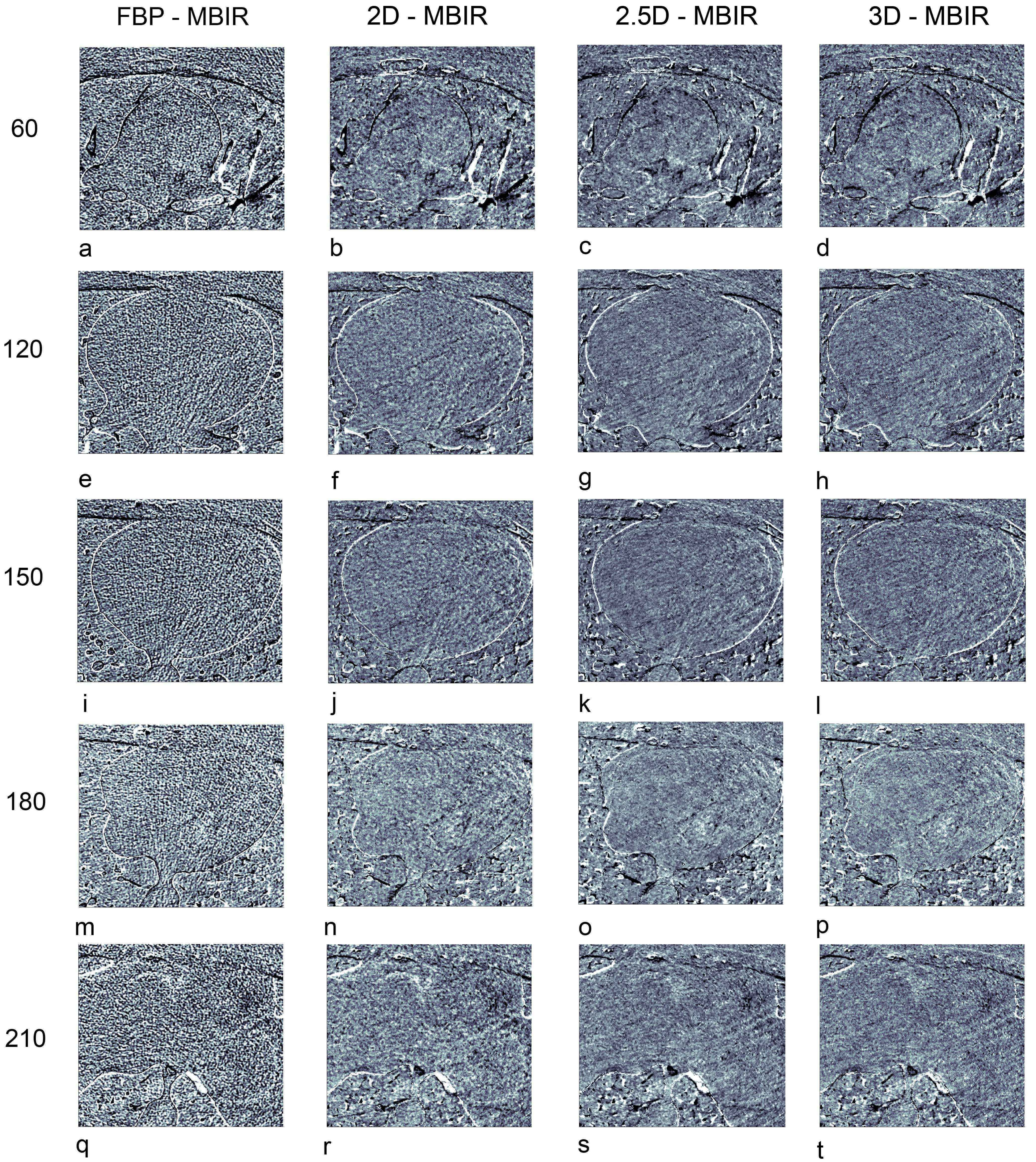}
\vspace{-0.5cm}
\caption{Comparison between difference images for selected slices from different regions of the reconstructed volume.
The slice numbers are: a-d) 60; e-h) 120; i-l) 150; m-p) 180; q-t) 210.
Columns from left to right: FBP, 2D, 2.5D and 3D DL-MBIR subtracted from MBIR.}
\label{fig10}
\end{figure}

Finally, the reconstructed slices are compared in Figures~\ref{fig7}, \ref{fig8}, \ref{fig9} and \ref{fig10}.
In Figures~\ref{fig7} and~\ref{fig8}, slice 90 from case study 1 and slice 150 from case study 2 were selected, respectively, for comparison between different methods.
Panels a through e in these Figures correspond to the reconstructed slice using MBIR, FBP, 2D, 2.5D and 3D DL-MBIR.
The difference between each reconstructed image and the MBIR result is plotted in panels f through i for FBP, 2D, 2.5D and 3D DL-MBIR, respectively.
All the three methods enhance the reconstruction in comparison with the FBP results. 
The PSNR values for the two cases shown in Figures~\ref{fig7} and~\ref{fig8} are provided in Table~\ref{Tab3}.
In addition, Figure 7 demonstrates that the 2.5D and 3D methods produce similar results while they diminish the streaking artifact originated from FBP input, which is present in the results obtained from 2D DL-MBIR (Figure~\ref{fig7}c and g).
The better performance of the 2.5D and 3D methods is also evident in Figure 8 where the obtained results are less noisy compared to the 2D DL-MBIR result.
The PSNR values in Table~\ref{Tab3} further corroborate these qualitative observations and the enhancements obtained by 2.5D and 3D DL-MBIR methods.

More slices, from case study 1, are plotted in Figures~\ref{fig9} and \ref{fig10}. In Figure~\ref{fig9}, each column corresponds to a different reconstruction method (MBIR, FBP, 2D, 2.5D and 3D DL-MBIR) while each row belongs to a separate slice number from the reconstruction volume (60,120,150,180, 210). 
The latter correspond to different organs in the reconstruction volume.
Assuming MBIR as the ground truth, we plotted the difference images between each method and MBIR in Figure~\ref{fig10}.
Analysis of the panels in Figures~\ref{fig9} and \ref{fig10} further demonstrates the promising prospect of using the proposed 2.5D method.
For examples in slices 60 to 150, while some streaking artifacts, that are originated from the FBP images, were still present in the results obtained by 2D DL-MBIR,  they are removed/reduced in the proposed 2.5D DL-MBIR akin to 3D DL-MBIR.
In addition, the similar performance of  2.5D and 3D DL-MBIR is evident in slices 180 to 210, where the results from both these networks have smoother reconstruction and have minimized the noise and artifacts compared to 2D DL-MBIR results.

\section{Conclusions}
\label{sec5}
In this paper, we proposed deep learning MBIR (DL-MBIR) as a fast algorithm for the approximate computation of MBIR images.
We developed 2D, 2.5D, and 3D DL-MBIR algorithms that uses both spatial and depth information from 3D volume of CT scans.
The 2.5D DL-MBIR algorithm offered image quality that was comparable or even better than full 3D processing with much less computation;
and 2.5D DL-MBIR also offered significantly better image quality than 2D processing with only slightly more computation.
Results were presented on medical CT images that demonstrated the potential of DL-MBIR to greatly improve image quality 
at minimal computational cost.

\bibliographystyle{IEEEbib}
\bibliography{Refs}

\end{document}